\documentclass[eps]{elsart}
\usepackage{epsf}

\def\jk {J_{K}}
\def\jh {J_{H}}

\def\sig {\sigma}

\def\om {\omega}

\def\ek {\varepsilon_{k}}
\def\be {\begin{equation}}
\def\eq {\end{equation}}

\begin{document}

\begin{frontmatter}
 \title{Short-range antiferromagnetic correlations in Kondo insulators} 
 \author{Tatiana G. Rappoport, M. S. Figueira and M. A. Continentino}
 \address{Instituto de F\'{\i}sica, Universidade Federal Fluminense Av.
    Litor\^anea, s/n$^o$ - 24210-340 Niter\'oi, RJ, Brazil}

\begin{abstract} 
We study the influence of short range antiferromagnetic correlations between 
local $f$-electrons on the transport and thermodynamic properties of Kondo 
insulators. We use a Kondo lattice model with an additional Heisenberg
interaction between nearest neighboring f-moments, as first proposed  
by Coqblin {\it et al.}\cite{coqblin1}  
for metallic heavy fermions. The inter-site magnetic correlations  
produce an effective bandwidth for the $f$-electrons. They are treated  
on the same footing as the local Kondo correlations such that  
two energy scales appear in our approach. 
We discuss the competition between these two scales on the physical 
properties. 
\end{abstract} 


\end{frontmatter}

\section{Introduction}

Kondo insulators form a group of compounds that at high temperatures behave
as dirty metals but at very low temperatures have their thermodynamic and
transport properties determined by the existence of a small gap (10-100K)
that arises from the hybridization of local electrons and the conduction
band~\cite{fisk1,fisk2}. This family of compounds can be characterized as
strongly correlated semiconductors due to the $f$ or $d$ character of the
relevant electrons and includes $FeSi$~\cite{c0}, $Ce_{3}Bi_{4}Pt_{3} $~\cite
{c1a,c1b,c1c}, $SmB_{6}$~\cite{c2}, $YbB_{12}$~\cite{c3} and $CeFe_{4}P_{12}$%
~ \cite{c4}.

Many theoretical models have been used to describe Kondo insulators. Some of
them considered a two-band system, a large uncorrelated band of $s$
electrons and another narrow correlated band which describes the $f$ or $d$
electrons~\cite{mucio1,mucio2,mucio3}. Others considered an Anderson
lattice, a conduction band and localized $f$ states, with the correlation in
these states generally treated within the slave-boson method~\cite
{slave1,slave2}.

Although Kondo insulators do not present long range magnetic order, the
occurrence of short range antiferromagnetic correlations in these materials
has been observed by inelastic neutron scattering \cite{anti}, suggesting
that a good model to describe Kondo insulators must include
antiferromagnetic correlations between neighbors $f$-moments. We also point
out that the magnetic susceptibility of Kondo insulators in the Curie-Weiss
regime has a negative Curie temperature, as shown in Table 1 for some
compounds. This is a clear indication of the presence of antiferromagnetic
correlations in these materials.

\begin{table}[bbp]
\centering 
\par
\begin{tabular}{|l|l|}
\hline\hline
Compound & $~~~~~\Theta$ \\ \hline
$Ce_{3}Bi_{4}Pt_{3}$ & - 125 K ~\cite{c1c} \\ 
$YbB_{12}$ & -79 K ~\cite{japon} \\ 
$FeSi$ & -1030 K ~\cite{c0} \\ \hline\hline
\end{tabular}
\caption{Table of Curie Temperature for some compounds}
\end{table}

In this letter we study some physical properties of the Kondo lattice taking
into account short-range antiferromagnetic correlations between the
localized $f$-electrons (only first neighbors) as first proposed by Coqblin 
\textit{et al.}\cite{coqblin1}. These authors have used this approach to
describe metallic heavy-fermions while here we shall use it to investigate
Kondo insulators. We calculate the density of states and study its behavior
with increasing temperature and increasing magnetic correlations. We
evaluate some physical properties such as optical conductivity, magnetic
susceptibility and electrical resistivity. The format of the present letter
is as follows. In section \ref{s2} we present the Hamiltonian and describe
the approach used. In section \ref{s3} we calculate the density of states
and some physical quantities in order to analyze the influence of magnetic
correlations in the system.

\section{Model Hamiltonian\label{s2}}

We consider the following Hamiltonian to describe the system \cite{coqblin1}:

\begin{equation}
H=\sum_{k,\sigma }{\varepsilon _{k}n_{k\sigma }^{c}} + E_{0}\sum_{i,\sigma } 
{n_{i\sigma }^{f}} - {J_{K}} \sum_{i}{\vec{s}_{i}\cdot{\vec{S}_{i}}} - {J_{H}%
}{\ \sum_{i,\delta }} {\vec{S}_{i}}\cdot{\vec{S}_{i+\delta}},  \label{hini}
\end{equation}
where $n_{k\sigma }^{c}=c_{k\sigma }^{\dagger }c_{k\sigma }$, $n_{i\sigma
}^{f}=f_{i\sigma }^{\dagger }f_{i\sigma }$. $\vec{S_{i}}$ are the spin
operators associated to localized $f$-moments and $\vec{s_{i}}$ to
conduction electrons at site $i$. The first term represents the conduction
band. The operators $c_{k\sigma }^{\dagger }$ and $c_{k\sigma }$,
respectively, create and annihilate electrons in the conduction band state
labeled by the wave vector $\vec{k}$ and spin $\sigma$. The second term,
proportional to $E_{0}$, represents the binding energy of the $f$-electrons.
The third term is the $s-f$ exchange interaction $J_{K}$ which gives rise to
the Kondo coupling. In this model, following Coqblin \textit{et al.} \cite
{coqblin1}, we add a Heisenberg-like interaction $J_{H}$ between nearest
neighboring $f$-moments.

We examine the model above, specifically in the situation where the ground
state of the system is an insulating state. First we construct an effective
Hamiltonian, rewriting the original one, Eq.(\ref{hini}), in terms of new
operators, as performed by Ruppenthal \textit{et. al}~\cite{rs}. Since we
intend to describe the Kondo effect, we choose an operator $\hat{\lambda}
_{i,\sigma }$ that couples $c$ and $f$ electrons. In order to describe
short-range magnetic correlations we introduce another operator $\hat{\Gamma}%
_{i,\sigma }$ that couples $f$ electrons in neighboring sites. These
Hermitian operators are given by: 
\begin{equation}
\hat{\lambda}_{i\sigma }\equiv \frac{1}{\sqrt{2}}(c_{i\sig}^{\dagger }
f_{i\sigma}+f_{i\sigma}^{\dagger }c_{i\sigma})
\end{equation}
and 
\begin{equation}
\hat{\Gamma}_{i+\delta,\sigma }\equiv \frac{1}{\sqrt{2}}(f_{i\sigma}^{%
\dagger } f_{i+\delta,\sigma}+f_{i+\delta,\sigma}^{\dagger }f_{i\sigma}).
\end{equation}

The spin components can be written in terms of the new operators. Using an
additional constraint, that excludes the double $f$-occupancy, we get: 
\begin{eqnarray}
s_{i}^{x}S_{i}^{x}+s_{i}^{y}S_{i}^{y} = -\frac{1}{2}\sum_{\sigma}{\hat{%
\lambda} _{i\sigma }\hat{\lambda}_{i,-\sigma }},  \nonumber \\
s_{i}^{z}S_{i}^{z} =\frac{1}{4}({n_{i}^{c}}+{n_{i}^{f}})- \frac{1}{4}{%
n_{i}^{c}}{n_{i}^{f}}-\frac{1}{2}\sum_{\sigma}{\ \hat{\lambda}_{i,\sigma
}^{2}},  \nonumber \\
S_{i}^{x}S_{i+\delta }^{x}+S_{i}^{y}S_{i+\delta }^{y} = -\frac{1}{2}\sum_{
\sigma}{\hat{\Gamma}_{i+\delta ,\sigma }{\hat{\Gamma}_{i+\delta ,-\sigma }}},
\\
S_{i}^{z}S_{i+\delta }^{z} =\frac{1}{4}({n_{i}^{f}}+{n_{i+\delta}^{f}}) -%
\frac{1}{4}{n_{i}^{f}}{n_{i+\delta}^{f}}- \frac{1}{2}\sum_{\sigma}{\hat{%
\Gamma}_{i+\delta ,\sigma }^{2}},  \nonumber
\end{eqnarray}
where $n^{\alpha}_{i}=n^{\alpha}_{i\uparrow}+n^{\alpha}_{i\downarrow}$.
Since we are describing an insulating state at low temperatures, we use the
constraints $<n_{i}^{f}>=<n_{i}^{c}>=1$. Furthermore applying the
decoupling, 
\begin{equation}
<n_{i\sigma}^{\alpha }n_{j\sig}^{\beta }>\approx <n_{i\sigma}^{\alpha }>n_{j%
\sig }^{\beta }+<n_{j\sig}^{\beta }>n_{i\sigma}^{\alpha },
\end{equation}
we find that the terms containing the number operators cancel out. Note that
we are interested in non-magnetic solutions such that $<n_{i\sigma}^{\alpha
}>=<n_{i-\sigma}^{\alpha }>$. Finally we can write the Kondo and Heisenberg
parts of the Hamiltonian as: 
\begin{equation}
H_{K}=\frac{1}{2}J_{K}\sum_{i\sigma}{(\hat{\lambda}_{i\sigma }+\ \hat{\lambda%
} _{i,-\sigma })\hat{\lambda}_{i\sigma }},  \label{lam}
\end{equation}
\begin{equation}
H_{H}=\frac{1}{2}J_{H}\sum_{i+\delta \sigma}{(\hat{\Gamma}_{i+\delta,\sigma
} +\hat{\Gamma}_{i+\delta ,-\sigma})\hat{\Gamma}_{i+\delta ,\sigma}}.
\label{gam}
\end{equation}

Consistently with the decoupling used above, Eq.(5), we deal with the
product of operators in equations (\ref{lam}) and (\ref{gam}) introducing
the mean fields $\lambda _{i}=<\hat{\lambda}_{i,\sigma}>$ and $\Gamma _{i}=<%
\hat{\Gamma} _{i+\delta ,\sigma}>$. Furthermore due to translational
invariance, $\lambda _{i}=\lambda $ and $\Gamma _{i}=\Gamma $. Performing a
Fourier transform, using $\varepsilon _{k}=-\frac{W}{z}\sum_{R}{\cos {(kR)}}$
where $z$ is the number of neighbors and $2W$ is the band width, we obtain
the effective Hamiltonian: 
\begin{eqnarray}
H^{\prime}=\sum_{k,\sigma}{\varepsilon_{k\sig}n_{k\sigma}^{c}} +J_{K}\lambda
\sum_{k\sigma}{(c^{\dagger}_{k\sig}f_{k\sig}+ f^{\dagger}_{k\sig}c_{k\sig})}
+\sum_{k,\sigma}{(E_{0}-b\varepsilon_{k})n_{k\sigma}^{f}} \\
~~~~~~-2J_{K}\lambda^{2}-2J_{H}\Gamma^{2},  \nonumber  \label{hamiltonfinal}
\end{eqnarray}
where

\begin{equation}
b=\frac{J_{H}}{W}\Gamma.
\end{equation}

We have then arrived at a new picture where the magnetic correlations $J_{H}$
give rise to a narrow band of $f$-electrons of effective width $2bW$ (b is
temperature dependent). The interaction $J_{K}$ (using $J_{K}<0$, since we
are describing the Kondo effect) introduces an effective hybridization
between the conduction and the renormalized $f$ bands.

The new quasi-particles associated with the Hamiltonian (\ref{hamiltonfinal}%
) can now be obtained. This is done by calculating the one-electron Green's
functions which are found from their equations of motion. These Green's
functions are given by,

\begin{equation}
\mathit{G_{k,\sigma }^{ff}(\omega )}=\frac{A^{f}(\varepsilon _{k},\omega )}{%
\omega -\omega _{1}(\varepsilon _{k})}-\frac{A^{f}(\varepsilon _{k},\omega )%
}{\omega -\omega _{2}(\varepsilon _{k})},  \label{Gffb}
\end{equation}

\begin{equation}
\mathit{G^{cf}_{k,\sigma}(\omega)}=\frac{J_{K}\lambda}{(\omega_{1}
(\varepsilon_{k})-\omega_{2}(\varepsilon_{k}))}\biggl(\frac{1}{\omega
-\omega_{1}(\varepsilon_{k})}-\frac{1}{\omega-\omega_{2}(\varepsilon_{k})}%
\biggr),  \label{Gsfb}
\end{equation}

\begin{equation}
\mathit{G_{k,\sigma }^{cc}(\omega )}=\frac{A^{c}(\varepsilon _{k},\omega )}{%
\omega -\omega _{1}(\varepsilon _{k})}-\frac{A^{c}(\varepsilon _{k},\omega )%
}{\omega -\omega _{2}(\varepsilon _{k})}.  \label{Gssb}
\end{equation}
where $G_{k,k^{\prime }}^{\alpha \beta }=<<\alpha _{k\sig}^{\dagger }|\beta
_{k^{\prime }\sigma }>>$ for $\alpha ,\beta =c,,f$ and 
\begin{equation}
A^{c}(\varepsilon _{k},\omega )=\frac{\omega -(b\varepsilon _{k}+E_{0})}{%
\omega _{1}(\varepsilon _{k})-\omega _{2}(\varepsilon _{k})},
\end{equation}
\begin{equation}
A^{f}(\varepsilon _{k},\omega )=\frac{\omega -\varepsilon _{k}}{\omega
_{1}(\varepsilon _{k})-\omega _{2}(\varepsilon _{k})}.
\end{equation}
We have expressed the Green's functions in terms of simple poles
corresponding to two hybridized quasi-particle bands with dispersion
relations given by 
\begin{equation}
\omega _{1;2}(\varepsilon _{k})=\frac{1}{2}[(1+b)\varepsilon _{k}+E_{0}]\pm 
\frac{1}{2}\sqrt{[(1-b)\varepsilon _{k}-E_{0}]^{2}+4J_{K}^{2}\lambda ^{2}}.
\label{polos}
\end{equation}

\begin{figure}[!h]
\epsfxsize=2.7 truein \centerline{\epsfbox{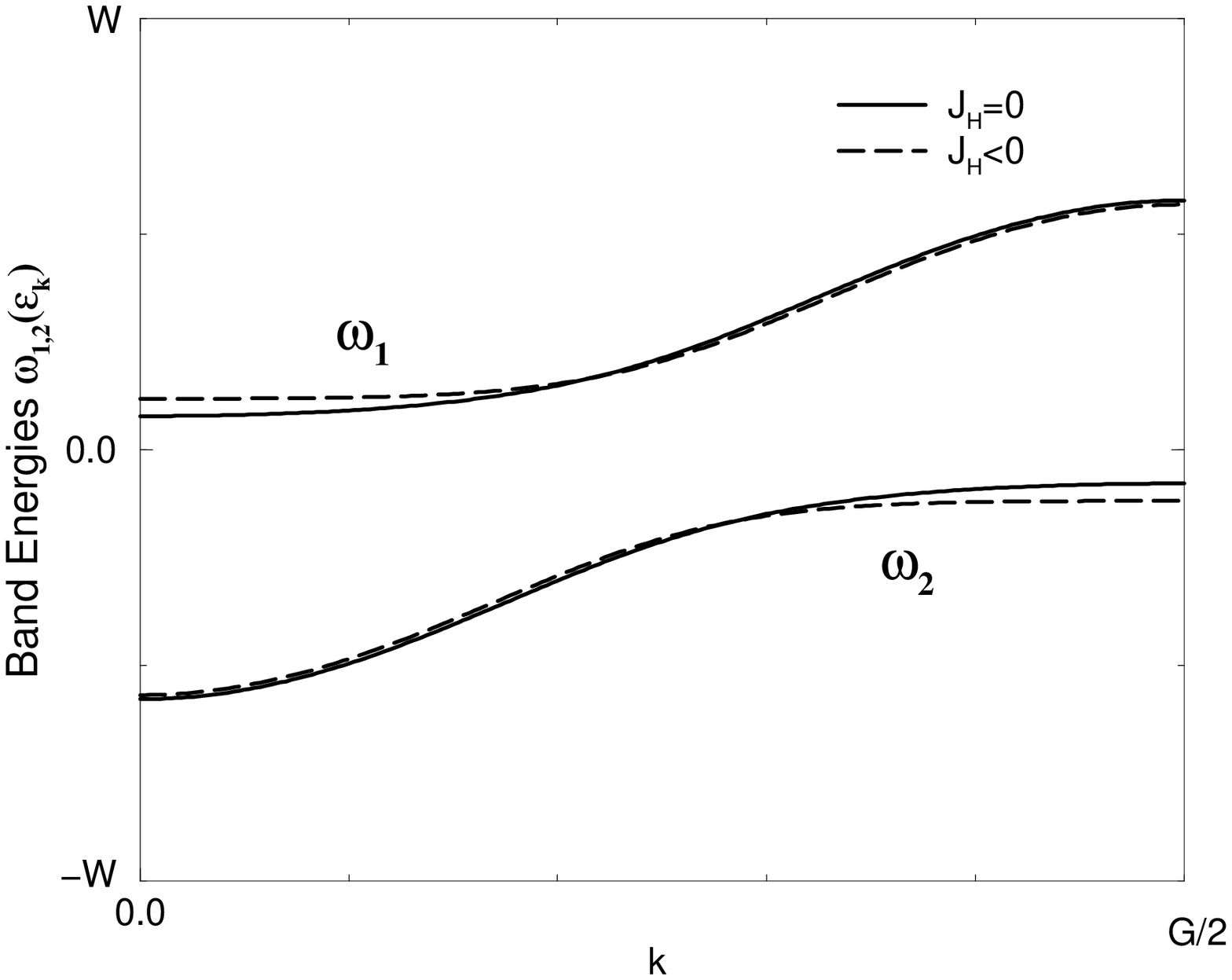}}
\caption{The energy dispersion relations $\omega_{1;2}(k)$ for the
hybridized bands found in the mean-field approximation using $\ek%
=-Wcos(ka/\pi) $, where $G/2=\pi/2a$ is half a reciprocal lattice vector.
Solid line is used for $\jh=0$ and dashed line for $\jh<0$.}
\label{fig1}
\end{figure}

As shown in figure~\ref{fig1} the two bands are separated by an indirect gap
between the lower band at $k=G/2$ and the upper band at $k=0$. For
increasing values of the magnetic correlation $|\jh|$, the indirect gap gets
larger while the direct gap remains constant. The direct gap $\Delta_{dir}$
is governed only by $\jk$. It is easy to see that $\Delta_{dir}=2\jk\lambda$
and it occurs at $E_{0}=(1-b)\ek$.

An insulating ground state, as observed in $Ce_{3}Bi_{4}Pt_{3}$\cite{c1a} or 
$YbB_{12}$\cite{c3}, is obtained when the lower band is completely filled.
This corresponds to the half-filled band case where the occupation numbers
assume the values $<n_{i}^{f}>=<n_{i}^{c}>=1$. Furthermore, we take $%
E_{0}=\mu =0$ and symmetric bands with respect to the chemical potential $%
\mu $. Since the $f$-moments are correlated antiferromagnetically, which
corresponds to $J_{H}<0$, for $J_{K}\neq 0$ we always have an insulating
ground state with the chemical potential fixed in the middle of the gap of
the density of states.

The parameters $\lambda$ and $\Gamma$ are obtained from the two coupled
equations that have to be solved self-consistently for each temperature: 
\begin{eqnarray}
1={J_{K}}\sum_{k}{\frac{1}{(\omega_{1}-\omega_{2})} \left(\frac{f(\omega_{1})%
}{\omega-\omega_{1}}-\frac{f(\omega_{2})} {\omega-\omega_{2}}\right)} \\
\Gamma=-\sum_{k}{\frac{1}{(\omega_{1}-\omega_{2})}\left (\frac{(\omega_{1}-%
\ek)f(\omega_{1})}{\omega-\omega_{1}}-\frac{(\omega_{2}-\ek)f(\omega_{2})} {%
\omega-\omega_{2}}\right)}
\end{eqnarray}
where $\mathit{f}(\omega)=\frac{1}{\exp{\beta(\omega-\mu)}+1}$ is the Fermi
distribution and $\omega_{1;2}=\omega_{1;2}(\ek)$ is given by Eq.(\ref{polos}%
)

The present method has many similarities with the slave-boson mean field
approach and like it, presents a critical temperature $T_{c}$ where the
conduction and the $f$ electrons become decoupled~\cite{slave1,slave2}. In
our model, this arises for $\lambda(T_{c})=\Gamma(T_{c})=0$ which implies
the vanishing of the effective hybridization term $\lambda$. So, these
mean-field methods are useful to study the properties of Kondo insulators
only in the low temperature regime, below the critical temperature $T_{c}$
of the spurious phase transition. We shall present results here for $T\ll
T_{c}$ such that the present approach is valid.

\section{Theoretical Results and Comparison with Experiments\label{s3}}

As discussed previously, the magnetic correlations modify the band structure
of the system, increasing the indirect gap. In this section we analyze the
influence of these correlations in some physical properties, calculating $%
\lambda$ and $\Gamma$ self-consistently for each value of temperature. For
simplicity we consider a uniform density of states of width $2W$ for the
conduction electrons.

\subsection{Density of states}

\begin{figure}[!h]
\epsfysize=2.7truein \centerline{\epsfbox{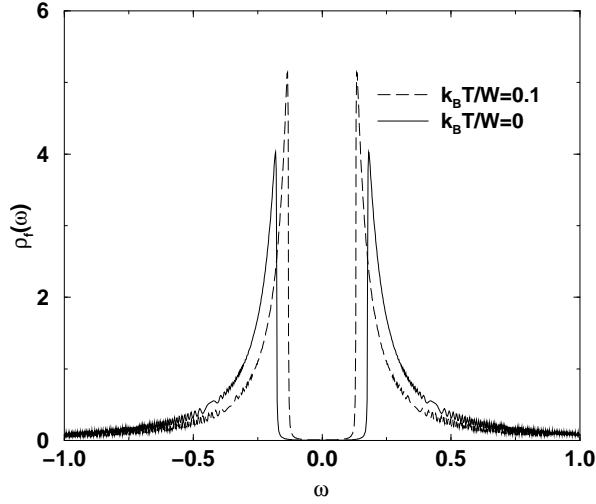}}
\caption{$f$ contribution to the density of states for different
temperatures. $J_{K}/W=-0.4$, $J_{H}/W=-0.1$.}
\label{denst}
\end{figure}

\begin{figure}[!h]
\epsfysize=2.7truein \centerline{\epsfbox{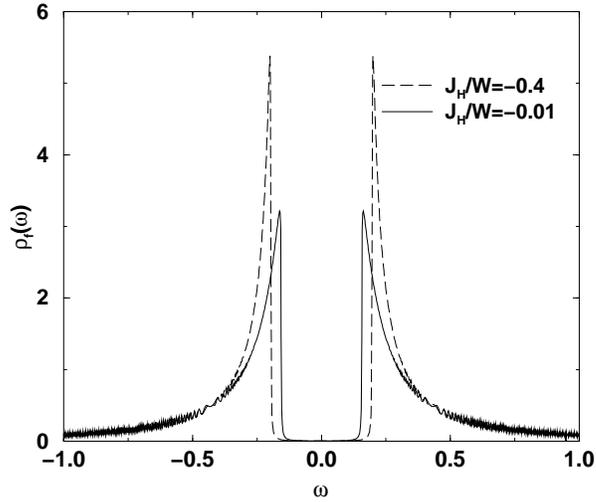}}
\caption{$f$ contribution to the density of states for two different values
of $J_{H}$ and $k_{B}T/W=0$.}
\label{densjh}
\end{figure}

Using that $<n^{\alpha,\beta}_{k,\sigma}>=\frac{1}{\pi} Im[G^{\alpha,%
\beta}_{k,\sigma}(\omega)]$ we obtain the $c$ and $f$ contributions to the
density of states. In figures \ref{denst} and \ref{densjh}, we show the
density of states which consists of two hybridized bands separated by a gap.
The density of states is very sharp near the band edges independently of the
form of the unperturbed bands.

Temperature renormalizes the density of states decreasing the gap (figure 
\ref{denst}). With increasing temperatures the density of states become more
peaked near the band edges. The density of states also depends on the
magnetic correlations represented here by the parameter $b$. The last term
in Eq.(\ref{hamiltonfinal}) produces an effective $f$-band and its weight
renormalizes with $b$. In a system where the Fermi level is inside the gap,
the main effects of short range antiferromagnetic correlations are to
increase the gap and enhance the density of states near the band edges, as
we can see in figure \ref{densjh}.

\subsection{Optical conductivity}

The optical conductivity can be written as~\cite{cond}: 
\begin{eqnarray}
\sigma(\nu)=\frac{\pi}{2W}\sum_{\sigma}{\int{d\varepsilon\int{d\om %
\rho^{c}_{\sigma}(\varepsilon,\omega)\rho^{c}_{\sigma}(
\varepsilon,\omega+\nu)}}} {{\frac{[f(\omega)-f(\omega+\nu)]}{\nu}}}
\end{eqnarray}
where $\rho^{c}_{\sigma}(\varepsilon,\omega)$ is the one particle spectral
density of the conduction electrons.

\begin{figure}[h]
\epsfysize=3.0truein \centerline{\epsfbox{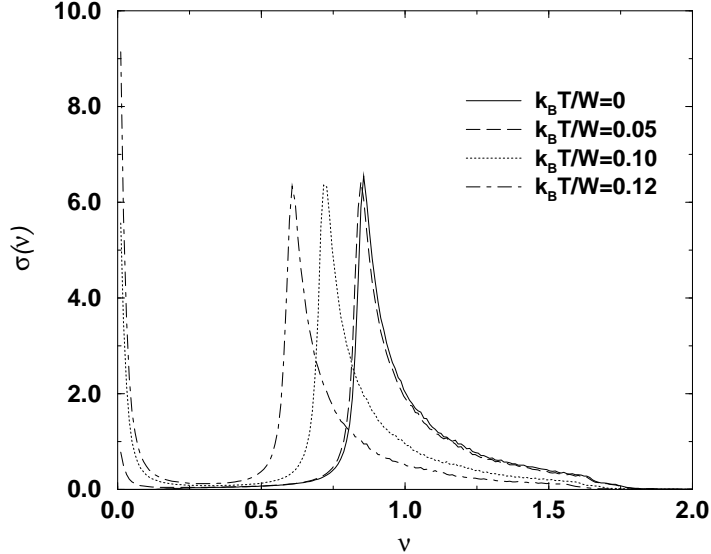}}
\caption{Optical conductivity for $\jk/W=-1.0$, $\jh/W=-0.1$ varying the
temperature. The frequency h$\nu $ is in units of the bandwidth $W$.}
\label{ocond}
\end{figure}

In the optical conductivity, shown in figure \ref{ocond}, the behavior of
the gap involves two characteristic temperature scales as observed
experimentally for $Ce_{3}Bi_{4}Pi_{3}$~\cite{ocondce} and $FeSi$~\cite{c0}.
Firstly the high temperature regime, compared to the width of the
conductivity gap $\Delta_c$, where the gap $\Delta_{c}$ itself is strongly
renormalized with $T$, as we can see in figure \ref{ocond}. The optical
conductivity also presents a Drude-like feature at low frequencies, as
observed recently by Gorshunov \textit{et al}~\cite{ocon3} in dynamical
conductivity measurements in $SmB_{6}$. This peak indicates that the
low-energy transport is determined by free charge carriers. Then the second
temperature scale is set by the width of this Drude peak. Only at small
temperatures the gap is completely opened. At these low temperatures the
variation in the intensity of this peak, whose weight is transferred to the
high frequency region of the spectrum, is the main effect of temperature.
The gap itself remaining unchanged as the low-energy behavior changes from
being dominated by free carriers to more localized carriers. We point out
that the interaction $\jh$ has no effect in the gap size of the optical
conductivity.

\subsection{Magnetic susceptibility and electrical resistivity}

In this subsection we calculate the electrical resistivity and the magnetic
susceptibility for different values of $J_{H}$ to analyze the influence of
antiferromagnetic correlations in these properties. The electric
conductivity is obtained from the limit $\nu \rightarrow 0$ of the optical
conductivity. 
\begin{equation}
\sigma (0)=\frac{\pi \beta }{2W}\int {d\varepsilon \int {d\omega [\rho
_{\sigma }^{c}\ (\varepsilon ,\omega )]^{2}f(\omega )[1-f(\omega )]}},
\label{condu}
\end{equation}
where $\beta \equiv 1/T$. The resistivity is given by $\rho =1/\sigma (0)$.
Notice that this expression must be used with care. In the problem
considered here, there is translational invariance and consequently\textbf{\
k }is a good quantum number. However in real systems  impurity scattering is
always present and this limits the electron mean free path. In the
calculations presented below, this is taken into account by including a
finite lifetime for the conduction electrons. Formally this is done by
replacing,  $\omega \rightarrow \omega +i\Gamma $, in the Green 's
functions, where $\Gamma $ is temperature independent. We point out that
here we are interested in the temperature dependence of the conductivity
which is not affected by the magnitude of $\Gamma $.

If an external magnetic field $h$ with its direction along the z-axis is
applied, the quasi-particle energies become 
\begin{equation}
\omega _{i,\sigma }(\varepsilon _{k})=\omega _{i}(\varepsilon _{k})-\sigma h.
\label{eh}
\end{equation}
The free energy in the presence of the field, using the new quasi-particles
energies (\ref{eh}), is written as 
\begin{equation}
F(\lambda ,\Gamma ,T)=-\frac{1}{\beta }\sum_{i=1,2}\sum_{k,\sigma }{{\ln {%
\biggl(1+\exp {({-\beta \omega _{i,\sigma }(\varepsilon _{k}))}\biggr) }}}}%
+2J_{H}\lambda ^{2}+2J_{H}\Gamma ^{2}  \label{free}
\end{equation}
and the magnetic susceptibility is given by 
\begin{equation}
\chi (T)=\frac{\partial ^{2}F}{\partial h^{2}}_{h=0}=\frac{1}{4}\beta
\sum_{i=1,2}\sum_{k,\sigma }{{sech^{2}\left( \frac{1}{2}\beta \omega
_{i,\sigma }(\varepsilon _{k})\right) }}.
\end{equation}
\begin{figure}[h]
\epsfysize=2.3truein \centerline{a)\epsfbox{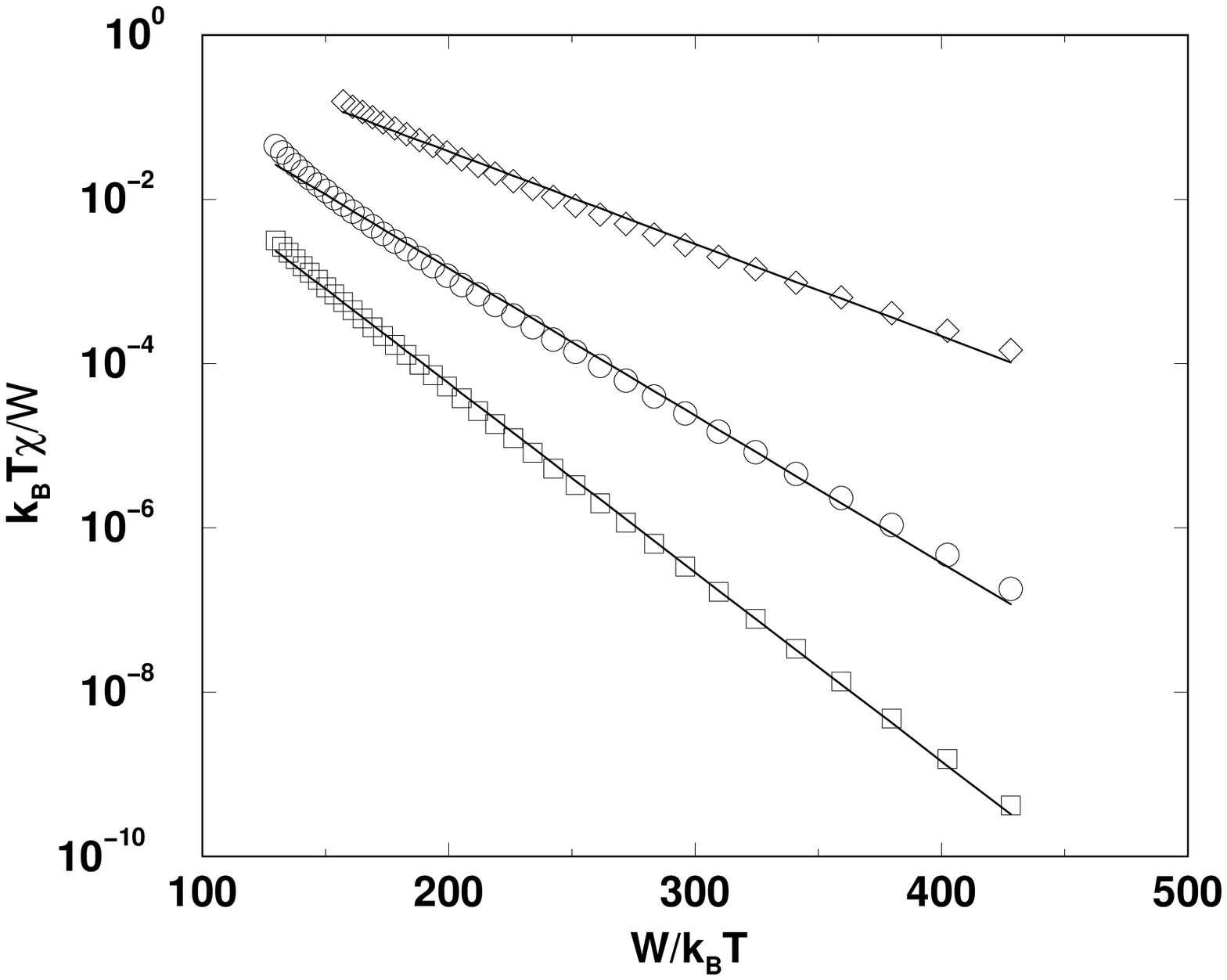} b)\epsfysize 
=2.3truein \epsfbox{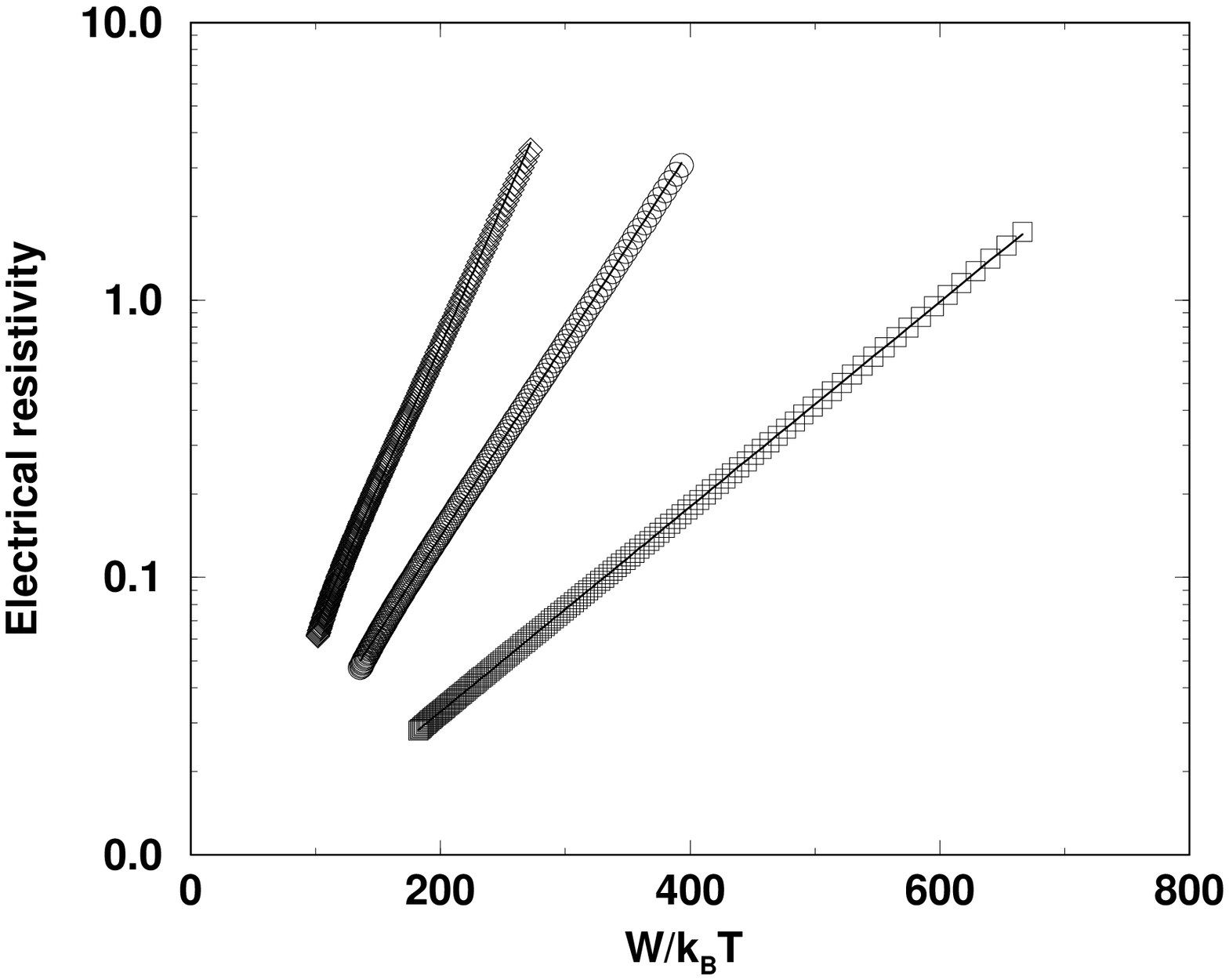}}
\caption{Temperature behavior of the calculated susceptibility (a) and
resistivity (b) showing activation behavior. The fitting using expressions
given in the text are shown as full lines for different values of $\jh$ and $
\jk 
/W=-0.4$.}
\label{ajustes}
\end{figure}

\begin{figure}[!h]
\epsfysize=3.0truein \centerline{\epsfbox{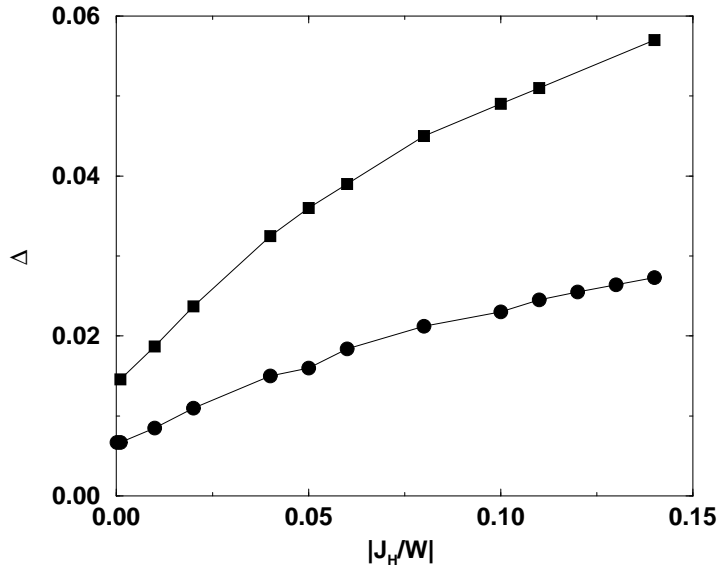}}
\caption{Magnetic (full squares) and transport (full circles) gaps versus $
\jh$ for $\jk/W=-0.4$.}
\label{gaps}
\end{figure}

The magnetic susceptibility and electrical resistivity can be obtained from
the expressions above for different values of $\jh$ and $\jk$ fixed. The
calculated low temperature resistivity and susceptibility curves were fitted
using the activated forms $\chi (T)=(C/T)\exp {(-\Delta _{\chi }/kT)}$ and $%
\rho (T)=\rho _{0}\exp {(-\Delta _{\rho }/kT)}$, which describe the
experimental data in Kondo insulators. As we can see in figure \ref{ajustes}%
, these analytical expressions give also a good description of the
theoretical results, indicating that our model describes very well the low
temperature properties of these systems. The fittings of the theoretical
results yield values for the transport (${\Delta _{\rho }}$) and magnetic (${%
\Delta _{\chi }}$) gaps \cite{c0}. In figure \ref{gaps} we show these gaps
for different values of $\jh$. $\Delta _{\chi }$ and $\Delta _{\rho }$
increase for increasing values of $|\jh/W|$ but their ratio remains the same
($\Delta _{\chi }/\Delta _{\rho }\approx 2.2$) suggesting that magnetic
correlations do not influence the relation between them.

\subsection{Analysis}

Experimental results yield different values for the gaps of Kondo
insulators. The gap measured in optical conductivity $\Delta_{c}$ has larger
values than the gaps measured in neutron scattering and susceptibility
measurements,$\Delta_{\chi}$ or in transport measurements $\Delta_{\rho}$%
\cite{ocondce}. On the other hand, comparing the spin and transport gaps,
experimental results show that $\Delta_{\chi}\geq \Delta_{\rho}$~\cite
{c0,c1c}. The present theory accounts for these observations and introduces
a new discussion about the roll of short-range magnetic correlations in the
properties of these gaps.

We have discussed in section 2 the influence of antiferromagnetic
correlations in the indirect and direct gaps in the density of states.
Although in the calculations of section 3 we used a square band, without a
defined dispersion relation, we point out that, since our results show that
the frequency gap $\Delta _{c}$ in the optical conductivity does not
renormalize as $\jh$ varies, we may conclude that this quantity is
determined by the direct gap. This is expected, since due to the negligeable
momentum of the photon, optical excitations involve essentially an energy
transfer. On the other hand the results in figure (\ref{gaps}) show that the
transport and susceptibility gaps are strongly renormalized by magnetic
correlations indicating the indirect nature of the gap that determines these
physical properties.

\section{Conclusions}

In this work we have discussed a model for Kondo insulators which takes into
account the influence of short-range magnetic correlations in these systems.
We used a Kondo lattice with an additional Heisenberg term and we treated
the problem within a mean-field approach in order to obtain an effective
Hamiltonian where the additional term gives rise to an effective band width
for the $f$ electrons. Although the Kondo interaction leads to an indirect
coupling between the local electrons, we have added explicitly to the
Hamiltonian an interaction between them. In this way we can make a
straightforward analysis of the influence of these magnetic correlations on
the properties of the system.

We have obtained the density of states of the new quasi-particles and
observed that their energy bands are renormalized by temperature and
antiferromagnetic correlations. We have calculated the magnetic
susceptibility and electric resistivity to get the magnetic and transport
gaps which turn out to be strongly renormalized by magnetic correlations.
Calculating the optical conductivity we find a Drude-like peak at very low
frequencies, as observed recently in dynamical conductivity measurements of $%
SmB_{6}$\cite{ocon3}. However, the gap in the optical conductivity remains
the same for different values of the strength of magnetic correlations. This
gap is determined by the direct gap in the band structure, while the
magnetic and transport gaps arise mostly from the indirect gap. The
antiferromagnetic correlations change substantially the physical properties
related to the indirect gap but do not change the quantities related to the
direct one. These observations can produce an insight for future
experimental analysis.

This letter discusses explicitly the short-range magnetic correlations which
have not been explored previously in the theoretical study of Kondo
insulators. The model is able to reproduce their physical properties,
including their temperature dependence, and is in good qualitative agreement
with experimental results.

\section*{Acknowledgments}

The authors like to thank M. A. Gusm\~{ao} for useful discussions and
sending a pre-print of his work. This work was supported by The Brazilian
agencies Conselho Nacional de Desenvolvimento Cientifico (CNPq) and CAPES.


\end{document}